\def\papertitle{FoleySet: A Multi-Level Human-Annotated Foley Sound Dataset}
\def\paperauthorA{Sunshiyu Wang}
\def\paperauthorB{Alexander Lerch}

\documentclass[twoside,a4paper]{article}
\usepackage{etoolbox}

\usepackage{dafx26v2}

\usepackage{amsmath,amssymb,amsfonts}
\usepackage{array}
\usepackage[T1]{fontenc}
\usepackage[utf8]{inputenc}
\usepackage[english]{babel}
\usepackage{caption}
\usepackage{subfig}
\usepackage{booktabs}
\usepackage{color}
\usepackage{times}
\usepackage{tabularx}
\usepackage{ragged2e}
\input glyphtounicode
\newcolumntype{Y}{>{\RaggedRight\arraybackslash}X}
\ninept
\usepackage{microtype}
\usepackage{units}
\usepackage{paralist}
\usepackage{pgfplots}
\pgfplotsset{compat=1.18}
\usepackage{placeins}
\usepackage{hyphenat}

\newcounter{numauth}
\newcounter{listcnt}
\newcommand\authcnt[1]{\ifdefined#1 \stepcounter{numauth} \fi}

\newcommand\addauth[1]{
\ifdefined#1
\stepcounter{listcnt}
\ifnum \value{listcnt}<\value{numauth}
\appto\authorslist{, #1}
\else
\appto\authorslist{~and~#1}
\fi
\fi}

\authcnt{\paperauthorB}

\def\authorslist{\paperauthorA}
\addauth{\paperauthorB}

\usepackage[pdftex,
  pdftitle={\papertitle},
  pdfauthor={Sunshiyu Wang and Alexander Lerch},
  pdfsubject={Proceedings of the 29th International Conference on Digital Audio Effects (DAFx26)},
  colorlinks=false,
  bookmarksnumbered,
  pdfstartview=XYZ
]{hyperref}

\usepackage{graphicx}

\title{\papertitle}

\affiliation
{\paperauthorA\ and \paperauthorB}
{Music Informatics Group \\ Georgia Institute of Technology \\ Atlanta, USA\\
{\tt \href{mailto:swang3214@gatech.edu}{swang3214@gatech.edu},
\href{mailto:alexander.lerch@gatech.edu}{alexander.lerch@gatech.edu}}
}

\begin{document}
\ifpdf 
  \DeclareGraphicsExtensions{.png,.jpg,.pdf}
\else  
  \DeclareGraphicsExtensions{.eps}
\fi


\maketitle

\begin{abstract}
In audiovisual post-production, Foley refers to synchronous sound effects associated with human actions, such as footsteps, cloth rustle, and prop handling, that are recreated to match the on-screen movements and interactions of characters. These sounds are often recorded by professional Foley artists using physical props. This resource-intensive workflow has motivated data-driven research on Foley, including tasks such as classification, retrieval, and generation; however, high-quality annotated Foley datasets for training remain scarce. To address this gap, we present FoleySet, a publicly available Foley dataset of 10,000 audio clips annotated with a two-level Foley taxonomy. This dataset provides a standardized, Creative Commons–licensed resource for data-driven Foley classification, retrieval, and generation.

\end{abstract}
    
\section{Introduction}

Foley refers to the synchronized reproduction of everyday sound effects for audiovisual media. Examples include footsteps, cloth movement, and object interactions such as doors opening or coins jingling. Unlike purely synthetic sound effects, Foley is often created and recorded in studio settings. In this process, Foley artists use a variety of props to recreate the sounds of real-world actions and material interactions in sync with the visual scene. This makes the production process labor-intensive, costly, and time-consuming. Alternatively, existing Foley sound libraries provide pre-recorded Foley sounds, but require significant time for both finding a fitting sound and manually synchronizing it to the on-screen events. Recent advances in deep-learning-based audio methods have opened up new data-driven directions for automating and supporting Foley-related workflows. However, compared with broader audio research areas, Foley-specific data-driven work remains constrained by the lack of structured, high-quality datasets. Such a dataset is therefore needed to support not only analyzing and synthesizing Foley but also a broader range of data-driven approaches to modern Foley production. Accordingly, we introduce \textit{FoleySet}, a publicly available dataset of 10,000 human-annotated Foley clips organized using a two-level taxonomy.\footnote{FoleySet
is publicly available at \url{https://zenodo.org/records/20735877}.} Building on prior research and to clearly define the scope of the dataset, we define Foley as sounds  arising from human-related actions---human--material interactions and human-produced sounds---rather than non-human sources such as animal vocalizations.

\begin{table*}
\centering
\caption{Comparison of the proposed Foley dataset with existing popular audio datasets.}
\label{tab:dataset_comparison}
\setlength{\tabcolsep}{5pt}
\renewcommand{\arraystretch}{1.15}
\begin{tabular*}{\textwidth}{l@{\extracolsep{\fill}}l r r r l}
\hline
\textbf{Dataset} & \textbf{Source} & \textbf{Total Dur. (hrs)} & \textbf{Samples} & \textbf{Classes} & \textbf{Annotation Source} \\
\hline
ESC-50           & Freesound               & 2.8    & 2{,}000       & 50         & Human annotation \\
UrbanSound8K     & Freesound               & 8.8    & 8{,}732       & 10         & Human annotation \\
FSD50K           & Freesound               & 108.0  & 51{,}197      & 200        & Tags + human validation \\
AudioSet         & YouTube                 & 5{,}833 & 2{,}084{,}320 & 527        & Human-verified labels \\
DCASE2023 Task 7 & US8K / FSD50K / BBC SFX & 6.2    & 5{,}550       & 7          & Task-specific labels \\
FoleyBench & Internet video (CC) & $\sim$13 & 5{,}000 & UCS tax. & Automated pipeline \\
MINT & AudioCaps & 111.2 & 40{,}016 & 4 (11 sub) & Captions + narrative text \\
6KSFx & Synthetic & 8.3 & 6{,}000 & 30 & Procedurally generated \\
\hline
FoleySet (ours)    & Freesound               & 9.5    & 10{,}000      & 9 (73 sub) & Human annotation \\
\hline
\end{tabular*}
\end{table*}

Several studies have explored sound effect classification, taxonomy construction, and library indexing for film post-production \cite{Peeters:2020:SoundClassPostProd, Moffat:2017:UnsupTaxonomy, Ma:2022:RepLearning, Lin:2023:Soundify}. On the synthesis side, existing deep-learning research on Foley has primarily focused on optimizing model architectures and improving synthesis quality. For example, the DCASE 2023 Task 7 organized a Foley sound synthesis challenge that provided a standardized evaluation framework, baseline systems, and a dataset spanning seven Foley sound classes \cite{Choi:2023:FoleySynthesis}. Subsequent Foley generation systems, such as T-Foley and MambaFoley, built on the same dataset and further improved performance through advances in model architecture \cite{Chung:2024:TFoley,Colombo:2025:MambaFoley}. Foley’s strong relation to visual media has also resulted in cross-modal synthesis research, leading to Video-to-Audio models such as AutoFoley and Diff-Foley \cite{Ghose:2020:AutoFoley,Luo:2023:DiffFoley}. 

Together, these works demonstrate the growing technical pro\-gress toward data-driven Foley workflows, while also highlighting the need for Foley-specific datasets and taxonomies. Several public audio datasets contain relevant sound-effect categories, but their taxonomies are typically organized around broad semantic classes rather than the action- and material-level distinctions central to Foley practice \cite{fonseca2022fsd50k}. As a result, they do not provide a unified Foley-specific hierarchy or fine-grained annotations for distinguishing subtle differences in action, material, and synchronization context. This mismatch motivates the taxonomy and annotation design of FoleySet.

More specifically, \textit{FoleySet} contains 10,000 audio clips collected from Freesound and annotated using a two-level taxonomy comprising 9 major categories and 73 sub-categories. Alongside each audio file, we collected user-provided metadata from the corresponding Freesound webpage, including tags and textual descriptions \cite{Freesound}. All audio clips were normalized, trimmed to remove leading silence, and segmented into multiple shorter clips when longer than \unit[5]{s}. Each clip was also assigned a one-shot/multi-shot label to indicate whether it contains a single sound event or repeated occurrences. The released dataset includes only clips distributed under Creative Commons Zero licenses that permit reuse and redistribution. Taken together, FoleySet provides a standardized and reproducible resource for fine-grained Foley research.


The remainder of this paper is structured as follows. Section~2 reviews related audio datasets. Section~3 describes the development of our Foley-specific taxonomy. Section~4 details the dataset construction pipeline. Section~5 presents dataset statistics. Section~6 reports benchmark classification results, and Section~7 concludes the paper.

\section{Existing datasets}

This section reviews existing audio datasets that are relevant to FoleySet in terms of content, collection process, or labeling design. Their main characteristics are summarized in Table~\ref{tab:dataset_comparison}.

A closely related open Foley-focused dataset was released for the DCASE 2023 Task 7 ``Foley Sound Synthesis'' challenge\footnote{\url{https://dcase.community/challenge2023/task-foley-sound-synthesis}}. The dataset contains 5,550 audio excerpts (4,850 for development, 700 for evaluation) drawn from UrbanSound8K\cite{salamon2014urbansound8k}, FSD50K \cite{fonseca2022fsd50k}, and the BBC Sound Effects library \cite{BBCSoundEffects}, and is organized into seven categories: \texttt{DogBark}, \texttt{Footstep}, \texttt{GunShot}, \texttt{Keyboard}, \texttt{MovingMotorVehicle}, \texttt{Rain}, and \texttt{Sneeze/Cough}. For the challenge, all audio was converted to mono signals with 16-bit and \unit[22,050]{Hz}, and zero-padded or segmented to a fixed duration of \unit[4]{s}. The dataset has played an important role in supporting recent research on Foley sound synthesis. However, its broader applicability is limited by its seven-class structure, which does not capture the diversity and granularity needed for fine-grained Foley tasks. Other Foley-related resources address complementary tasks.
FoleyBench~\cite{Dixit:2026:FoleyBench} provides video--audio--caption
triplets for evaluating semantic and temporal alignment in video-to-audio
generation, MINT~\cite{Fu:2024:MINT} focuses on image- and
narrative-text--driven Foley audio planning and dubbing, and
6KSFx~\cite{Garcia:2025:6KSFx} provides synthetically generated sound
effects for procedural-audio research. These datasets differ from FoleySet
in modality, source material, annotation design, and primary task.

Beyond the Foley-specific datasets described above, there exist a number of datasets that contain Foley-related classes among others. AudioSet is one of the largest general-purpose audio datasets and among the first to explicitly target broad coverage of everyday sound \cite{Gemmeke:2017:AudioSet}. It consists of about 2.1 million \unit[10]{s} clips drawn from YouTube.com and spans 527 classes covering speech, music, and sound events. Its labels are organized using a large-scale hierarchical audio event ontology, and each clip can receive one or more labels that are validated by multiple human raters through majority voting. The underlying ontology of AudioSet comprises 632 categories organized in a hierarchy up to six levels deep. Several top-level branches — notably \texttt{Human sounds} and \texttt{Sounds of things} — contain classes that overlap with common Foley categories, such as \texttt{Footstep}, \texttt{Door}, \texttt{Glass}, \texttt{Hands}, and \texttt{Breathing}. However, these Foley-relevant classes are scattered across disparate branches of the ontology rather than consolidated under a unified Foley category, limiting its direct use as a standardized benchmark for fine-grained Foley research. Overall, AudioSet has been widely adopted for tasks such as audio representation learning, and has strongly influenced subsequent audio dataset development. That said, because the original audio is sourced from YouTube, AudioSet is not released as a fully open collection of redistributable audio files. Access to the underlying audio depends on third-party hosting, and clips may become unavailable over time due to deletion, licensing changes, or copyright-related restrictions, which reduces the long-term stability of the dataset.

To address some of AudioSet’s limitations, FSD50K was introduced as a sound event dataset based on Freesound audio with distributable waveform files. It contains 51,197 audio clips totaling 108.3 hours of multi-label audio, and is manually annotated using 200 classes drawn from the AudioSet Ontology. These 200 classes are hierarchically organized as a subset of the ontology, including 144 leaf nodes and 56 intermediate nodes. The class selection was driven primarily by the availability of suitable audio material on Freesound, with a focus on sound events produced by physical sound sources and production mechanisms. FSD50K includes classes that overlap with common Foley categories — such as \textit{Footsteps}, \textit{Door}, \textit{Knock}, and \textit{Glass} — but, as it adopts the AudioSet ontology, these classes are not organized under a unified Foley-specific taxonomy. Compared with AudioSet, FSD50K provides a fully open and redistributable benchmark for sound event research, with annotations obtained through a hybrid pipeline in which candidate labels are first inferred from Freesound tags and then confirmed through human validation.

\begin{figure*} 
  \centering
  \hspace*{-10mm}\includegraphics[width=\textwidth, trim=0 30 0 30, clip]{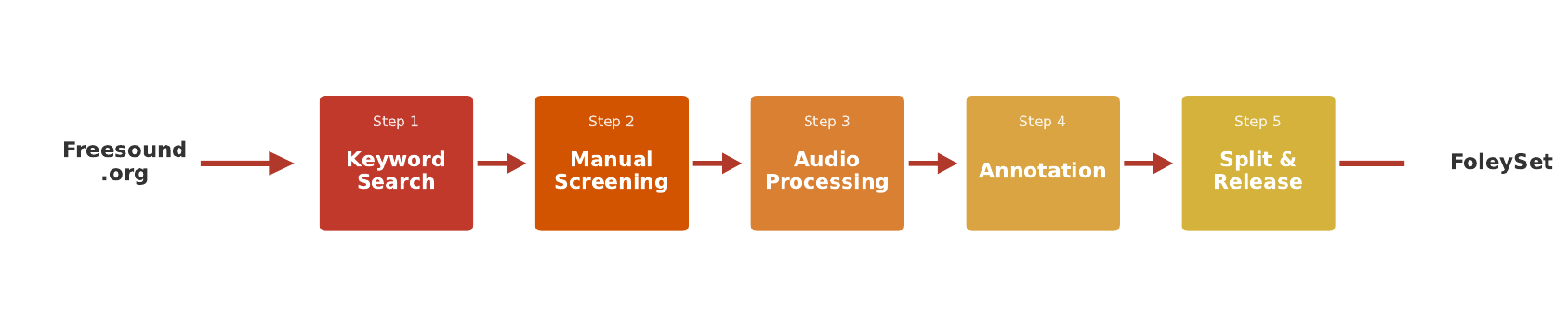}
  \caption{Overview of the FoleySet construction pipeline.}
  \label{fig:pipeline}
\end{figure*}

Two other datasets, ESC-50 \cite{piczak2015dataset} and UrbanSound8K \cite{salamon2014urbansound8k}, are also sourced from Freesound, similar to FSD50K. Both focus on more specific sound domains, which aligns with the goal of our work. ESC-50 is a fully human-labeled environmental sound dataset containing 2,000 \unit[5]{s} clips evenly balanced across 50 classes. These classes are organized into five broad categories: 
\texttt{animal sounds}, 
\texttt{natural soundscapes}, 
\texttt{ human non-speech sounds}, 
\texttt{interior/ domestic sounds}, 
and \texttt{exterior/ urban noises}.
The authors state that the 50 classes were manually selected to provide balanced coverage of major types of environmental sounds. They also note that the dataset design took into account the availability and diversity of Freesound recordings, as well as the usefulness and distinctiveness of each class. 
Several ESC-50 classes overlap with Foley-related sounds, particularly within the \texttt{human non\-/speech sounds} and \texttt{interior/ domestic sounds} categories.
However, as an environmental sound dataset, ESC-50 does not ensure that target sounds are isolated from background noise, nor does it indicate whether an event is one-shot or multi-shot, which is often important for Foley sounds.

UrbanSound8K is another dataset sourced from Freesound, designed specifically for urban sound classification and built upon a self-developed taxonomy tailored to this domain. To construct the labeling system, the authors first proposed an urban sound taxonomy consisting of four top-level groups ---\texttt{human}, \texttt{nature}, \texttt{mechanical}, and \texttt{music}--- along with a set of fine-grained and unambiguous leaf categories. This taxonomy was informed in part by sound sources frequently appearing in NYC 311 noise-complaint records. From this broader taxonomy, 10 classes were selected for UrbanSound8K: \texttt{air conditioner}, \texttt{car horn}, \texttt{children playing}, \texttt{dog bark}, \texttt{drilling}, \texttt{engine idling}, \texttt{ gun shot}, \texttt{jackhammer}, \texttt{siren}, and \texttt{street music}. The resulting dataset contains 8,732 clips, each with a maximum duration of \unit[4]{s}, for a total of approximately 8.75 hours of audio. To reduce class imbalance, no class contains more than 1,000 clips. UrbanSound8K demonstrates the value of combining domain-specific taxonomy design with publicly sourced audio data, which provide a useful reference for our Foley taxonomy design. Unlike FoleySet, UrbanSound8K targets urban environmental sound recognition and includes only a small number of categories directly related to human-performed Foley events.

Taken together, these prior datasets demonstrate both the usefulness and value of large-scale, broad-coverage audio datasets with weak or machine-assisted labels, useful for training a wide range of audio-related models, and the importance of domain-specific audio collections with carefully designed human annotations. Similar to ESC-50 and UrbanSound8K, the presented dataset FoleySet focuses on a specific sound domain rather than broad audio coverage. This narrow scope enables fine-grained and precise human labeling, targeting Foley-related research tasks that are not well served by large-scale weakly labeled datasets. 

\section{Taxonomy}
Foley is a commonly used term in audiovisual sound post-production; however, there is no universally accepted definition of its scope, and its boundary with the broader domain of sound effects remains blurred. Existing audio taxonomies therefore show overlaps and inconsistencies in how Foley and other sound categories are labeled. 
For example, the widely adopted Universal Category System (UCS)\cite{ucs_category_list2024} includes a Foley category with five subcategories (\texttt{Cloth}, \texttt{Feet}, \texttt{Hands}, \texttt{Misc}, and \texttt{Props}). However, many signature Foley sounds discussed in Foley-focused research and practice, such as punches, breathing, kissing, or door open/close, are placed under other primary UCS categories \cite{Lewis:2015:VentriloquialActs}. Thus, UCS provides a useful general sound-effects taxonomy but does not offer consistent or exhaustive coverage of Foley sounds.

Given the lack of established standards for structuring and annotating Foley sounds, we propose an operational and reproducible Foley taxonomy and describe its development process in this section. In a first step, we investigate the core characteristics of Foley based on prior research, artist interviews, industry reports, and books. Drawing on these  sources \cite{Lewis:2015:VentriloquialActs,Trento:2011:FoleySoundsVsRealSounds,Donaldson:2014:InvisibleBody}, we summarize several recurring characteristics of Foley sounds:
\begin{compactenum}[(i)]
\item Foley is a post-production practice of creating sound effects in synchrony with on-screen actions,
\item Foley chiefly relates to the imitation of human-related actions,
\item Foley commonly focuses on body-associated events and human interaction with surfaces and objects, and
\item Foley often supplies the subtle sonic details audiences expect to accompany visible actions on screen.
\end{compactenum} 
Based on these characteristics, we define Foley as sounds arising from human-related actions, including human-driven interactions with materials (e.g., glass clinks or metal impacts) as well as human-produced sounds (e.g., kissing or snoring). For the purposes of FoleySet, we therefore exclude non-human sources such as animal vocalizations (e.g., dog barks or wolf howls), while acknowledging that Foley may be used more broadly in professional production. This working definition provides the conceptual basis for the taxonomy construction process described below. 

To ensure the practical relevance of our taxonomy, we draw from reference data collected from commercial Foley sound libraries.
We analyzed the taxonomy structures and category distributions of seven commercial Foley libraries~\cite{soundideas_hollywood_foley_fx,soundideas_foley_sound_effects,soundideas_art_of_foley,soundideas_hd_foley,soundideas_ultimate_foley_collection,soundideas_fight_foley,soundideas_foley_footsteps} and extracted Foley-relevant keywords from each library’s metadata: category and sub-category labels, clip names, and descriptions. This allowed us to identify terms commonly used in commercial practice. These common keywords were then normalized into a unified vocabulary by resolving spelling variants and merging synonymous terms, yielding an initial pool of 315 candidate keywords. A subsequent manual review removed terms lacking clear sound or action references, and keywords otherwise falling outside our working definition of Foley. The remaining 119 keywords served as the primary inputs for taxonomy construction. Based on these keywords, we manually refined and consolidated them into 73 sub-categories which are organized into 9 major categories. The resulting two-level taxonomy, as well as the mapping from keyword to sub-category, is summarized in the Appendix in Table~\ref{tab:kw_sub_major_2col_vertical}.

\begin{figure*}[!t] 
  \centering
  \includegraphics[width=\textwidth]{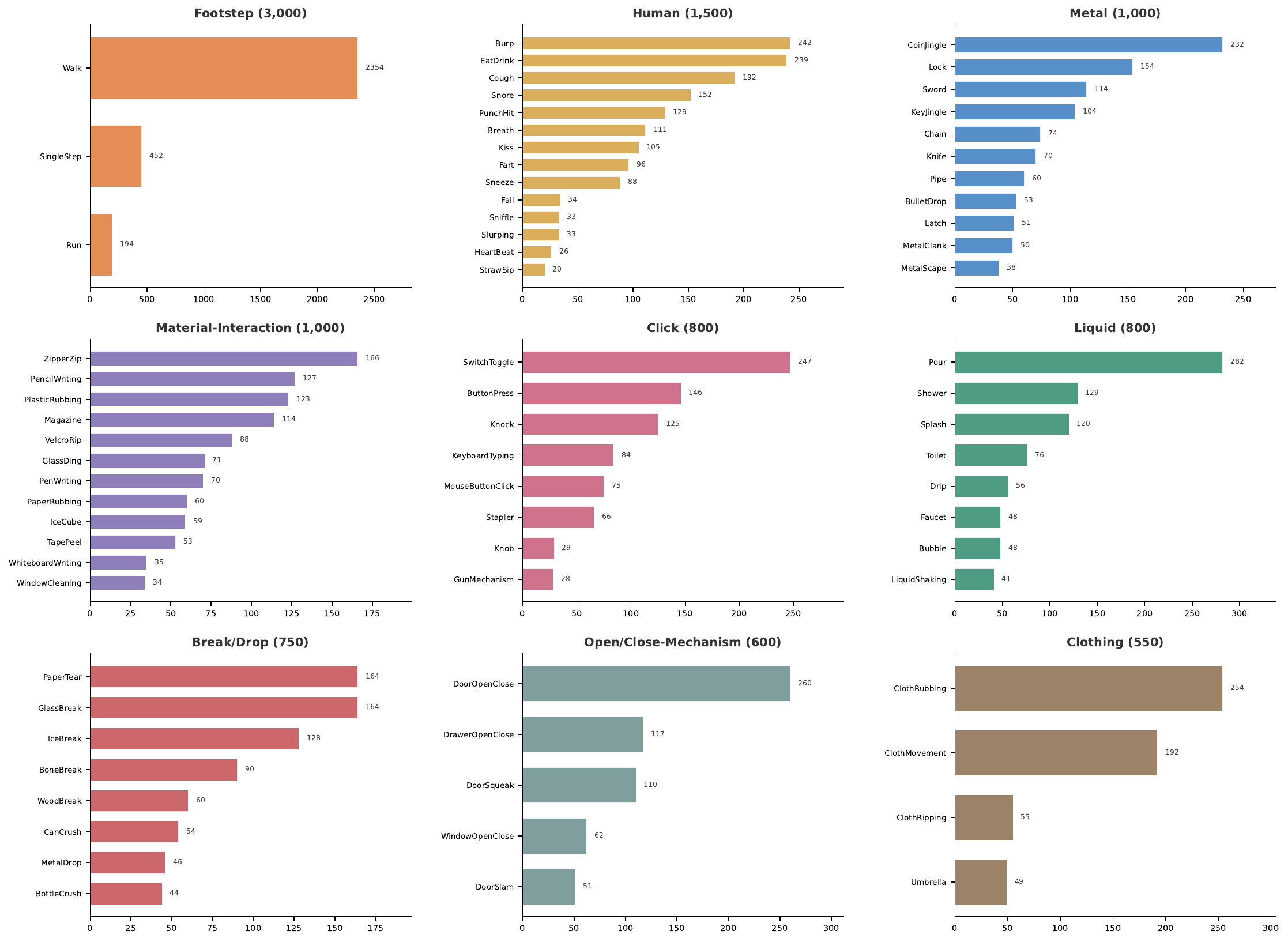}
  \vspace{-4mm}
  \caption{Sub-category distribution within each major-category, sorted by clip count. The number of total clips per category is indicated in parentheses.}
  \label{fig:sub_distribution}
  \vspace{-2mm}
\end{figure*}

\section{Dataset construction}

FoleySet is constructed through a multi-step pipeline designed to ensure that each audio clip is precisely labeled according to our proposed taxonomy while maintaining high audio quality and a balanced category distribution. All category and subcategory labels were assigned by a single annotator following the predefined taxonomy and annotation criteria. The construction pipeline is shown in Fig.~\ref{fig:pipeline}.

At the first stage, we collected approximately 23,500 candidate Foley clips from Freesound.org by querying the platform with sub-category names and corresponding keyword-based tags (accessed September 2025). We restricted our search to Creative Commons Zero (CC0)-licensed WAV clips with sampling rates of \unit[44.1]{kHz}, \unit[48]{kHz}, or \unit[96]{kHz}. This restriction allows unrestricted redistribution, while the resulting dataset may reflect the sounds and recording conditions more commonly found in CC0 uploads on Freesound. In addition, because Freesound does not formally review the audio quality of uploaded content and its user-provided tags and descriptions may be subjective or inconsistent, Stage 2 involved an initial manual screening to remove corrupted files, silent or excessively noisy clips, and recordings unrelated to Foley categories (e.g., guitar loops or synth drones). After this step, approximately 18,000 clips remained.

The remaining audio clips were resampled to \unit[44.1]{kHz} in Stage~3, converted to 16-bit mono, and loudness-normalized to $-23$~LUFS \cite{EBU:2011:R128} to ensure comparable levels and consistent format across the dataset. Each file was trimmed to start \unit[100]{ms} before the onset of the first audible event, providing a clean lead-in while removing unnecessary silence. To keep clip durations comparable, we enforced a maximum clip length of \unit[5]{s}: recordings longer than this threshold were segmented into multiple clips by cutting at the last silence point before the \unit[5]{s} limit, with each segment inheriting the original metadata. After segmentation, the pool contained approximately 15,000 clips.

In Stage 4, the remaining clips were manually reviewed and annotated. We set the following target clip counts per major category: Footstep (3,000), Human (1,500), Metal (1,000), Material-Interaction (1,000), Click (800), Liquid (800), Break/Drop (750), Open/Close-Mechanism (600), and Clothing (550). These targets reflect both the relative importance of each category in Foley practice and the availability of suitable source material on Freesound. All targets were met. Candidate clips were manually reviewed and assigned both a major-category and a sub-category label until the target count for each category was reached. In addition, each clip was assigned a one-shot/multi-shot tag to indicate whether it contained a single sound event or multiple sound events. We also retained the original user-provided Freesound tags and textual descriptions as additional metadata for each clip. The raw tags were lightly cleaned through a combination of automatic and manual steps: all tags were lowercased, single-character entries and tags containing digits (e.g., \texttt{ntg4}, \texttt{zoom-h2}) were removed, and space-separated tag strings were split into individual tokens. Morphological variants were merged into canonical forms (e.g., \texttt{footsteps} $\rightarrow$ \texttt{footstep}, \texttt{walking} $\rightarrow$ \texttt{walk}, \texttt{clothing} $\rightarrow$ \texttt{cloth}), and duplicate tags within the same clip were removed.

Finally, in Stage 5, we shuffled the clip order and renamed all files 
sequentially (e.g., \texttt{00001.wav} to \texttt{10000.wav}). The 
dataset is split into 8,000 training, 1,000 validation, and 1,000 
test samples (80/10/10). To prevent data leakage, all segments 
derived from the same original Freesound recording are assigned to 
the same partition. Under this constraint, we optimized the split 
assignment so that the subcategory distribution across partitions 
matches the overall distribution as closely as possible.

\section{Dataset statistics}
\label{sec:dataset_statistics}

\begin{table}
\centering
\caption{Overview of the fields provided in the released dataset metadata file.}
\label{tab:dataset_csv_fields_short}
\setlength{\tabcolsep}{6pt}
\small
\begin{tabular*}{\columnwidth}{p{.2\columnwidth} p{.7\columnwidth}}
\toprule
\textbf{Field} & \textbf{Description (example)} \\
\midrule
name & Clip filename in the released split (e.g., 09001.wav). \\
split & Data split indicator (e.g., train/val/test). \\[5mm]
category & Major-category label (9-way; e.g., Liquid). \\
sub-category & Sub-category label (73-way; e.g., Pour). \\
Oneshot-Multi & Event density tag (e.g., One-shot vs.\ Multi-shot). \\[5mm]
freesound-id & Source Freesound ID for traceability \\[5mm]
url & Link to the original Freesound page. \\[5mm]
username & Freesound uploader username for provenance (e.g., Rico\_Casazza). \\
original-tags & Original user-provided Freesound tags. \\[5mm]
description & Original Freesound text description. \\
\bottomrule
\end{tabular*}
\end{table}

The released dataset contains 10,000 audio clips (\unit[44.1]{kHz}, \unit[16]{bit} mono, max.\ length \unit[5]{s}) totaling approximately 9.5 hours of audio. Clip durations range from \unit[0.3]{s} to \unit[5]{s}, with a mean of \unit[3.4]{s} and a median of \unit[4.3]{s}. The 10,000 clips originate from 5,739 unique Freesound recordings (1.7 clips per source on average), with 4,128 sources contributing only a single clip. The dataset spans 9 major categories and 73 subcategories. At the major-category level, class sizes range from 550 (Clothing) to 3,000 (Footstep), with an imbalance ratio of approximately 5:1. At the subcategory level, the distribution exhibits a more pronounced long tail: class sizes range from 20 (StrawSip) to 2,354 (Walk), with a mean of 137 and a median of 84 clips per class. Of the 73 subcategories, 16 contain fewer than 50 clips and 41 contain fewer than 100 clips. While these statistics describe class coverage and imbalance, intra-class variation was not explicitly quantified in the current dataset. The original Freesound user-provided tags yield a vocabulary of 4,035 unique tokens across the dataset. In addition, each clip is annotated with a one-shot or multi-shot tag indicating whether it contains a single or multiple sound events; 23.6\% of clips are labeled One-shot and 76.4\% Multi-shot.
The full category and subcategory distributions are shown in Fig.~\ref{fig:sub_distribution}.

\begin{table}[t]
\centering
\caption{Top 50 original-tags in FoleySet, ranked by frequency.}
\label{tab:top50_original_tags}
\small
\setlength{\tabcolsep}{8pt}
\renewcommand{\arraystretch}{1.05}
\begin{tabular}{r l r @{\hspace{10pt}}|@{\hspace{10pt}} r l r}
\toprule
\# & Tag & Count & \# & Tag & Count \\
\midrule
1  & footstep   & 2903 & 26 & leather   & 404 \\
2  & walk       & 2576 & 27 & sneakers  & 402 \\
3  & foley      & 1949 & 28 & glass     & 400 \\
4  & step       & 1862 & 29 & soft      & 398 \\
5  & sound      & 1005 & 30 & click     & 391 \\
6  & running    & 982  & 31 & movement  & 386 \\
7  & foot       & 933  & 32 & human     & 378 \\
8  & water      & 793  & 33 & wooden    & 364 \\
9  & feet       & 774  & 34 & hit       & 344 \\
10 & wood       & 738  & 35 & ice       & 336 \\
11 & metal      & 676  & 36 & forest    & 335 \\
12 & paper      & 655  & 37 & male      & 328 \\
13 & open       & 635  & 38 & slow      & 325 \\
14 & door       & 630  & 39 & slide     & 323 \\
15 & gravel     & 591  & 40 & writing   & 310 \\
16 & floor      & 590  & 41 & rustling  & 307 \\
17 & cloth      & 573  & 42 & concrete  & 306 \\
18 & close      & 569  & 43 & switch    & 303 \\
19 & shoes      & 554  & 44 & shoe      & 301 \\
20 & boots      & 495  & 45 & break     & 301 \\
21 & run        & 477  & 46 & snow      & 293 \\
22 & dirt       & 448  & 47 & liquid    & 289 \\
23 & ground     & 434  & 48 & grass     & 282 \\
24 & crunch     & 429  & 49 & natural   & 279 \\
25 & fabric     & 410  & 50 & man       & 278 \\
\bottomrule
\end{tabular}
\end{table}

\begin{table}[!b]
\centering
\caption{Per-class results for 9-way major-category classification on the FoleySet test set (N=1000). P = Precision, R = Recall, Sup = Support (number of test samples).}
\label{tab:cls_perclass}
\small
\begin{tabular*}{\columnwidth}{l@{\extracolsep{\fill}} c c c r}
\toprule
\textbf{Class} & \textbf{P} & \textbf{R} & \textbf{F1} & \textbf{Sup} \\
\midrule
Brk/Drp      & 0.78 & 0.75 & 0.77 &  77 \\
Click        & 0.82 & 0.86 & 0.84 &  79 \\
Clothing     & 0.56 & 0.85 & 0.68 &  55 \\
Footstep     & 0.95 & 0.86 & 0.90 & 296 \\
Human        & 0.87 & 0.80 & 0.83 & 151 \\
Liquid       & 0.93 & 0.90 & 0.92 &  79 \\
Mat-Int      & 0.66 & 0.63 & 0.65 & 101 \\
Metal        & 0.78 & 0.87 & 0.82 & 103 \\
Op/Cl-Mech   & 0.80 & 0.86 & 0.83 &  59 \\
\bottomrule
\multicolumn{5}{l}{\footnotesize Brk/Drp = Break/Drop; Mat-Int = Material-Interaction;} \\
\multicolumn{5}{l}{\footnotesize Op/Cl-Mech = Open/Close-Mechanism; Wtd = Weighted.} \\
\end{tabular*}
\end{table}

Each clip's metadata is recorded as a single row in one CSV file containing all 10,000 entries. (Table~\ref{tab:dataset_csv_fields_short}). The file includes the major-category, sub-category, a one-shot/multi-shot event tag, and the original Freesound metadata: user-provided tags, textual description, uploader username, source ID, and URL. The user-provided tags were lightly cleaned and normalized by us; the 50 most frequent tags are shown in Table~\ref{tab:top50_original_tags}. These fields are intended to support not only supervised classification but also tasks such as text-based audio retrieval, captioning, and tag-based analysis. 

\begin{table}
\centering
\caption{Overall classification results on the Foley test set ($N = 1000$). Acc = Accuracy, P = Precision, R = Recall, F1 = F1-score, and w = weighted average.}
\label{tab:cls_overall}
\small
\begin{tabular*}{\columnwidth}{@{\extracolsep{\fill}}l c c@{}}
\toprule
& 9-way major-cat. & 73-way sub-cat. \\
\midrule
\textbf{Acc} & 0.82 & 0.64 \\
\textbf{P$_{\text{macro}}$} & 0.79 & 0.60\\
\textbf{R$_{\text{macro}}$} & 0.82 & 0.59\\
\textbf{F1$_{\text{macro}}$} & 0.80 & 0.56\\
\textbf{P$_{\text{w}}$} & 0.84 & 0.71\\
\textbf{R$_{\text{w}}$} & 0.82 & 0.64\\
\textbf{F1$_{\text{w}}$} &  0.83 & 0.64 \\
\bottomrule
\end{tabular*}
\end{table}

\section{Benchmark results}
We provide benchmark results for two classification tasks:
\begin{compactenum}[(i)]
	\item \textbf{Task 1}: 9-class major-category classification and 
	\item \textbf{Task 2}: 73-class sub-category classification. 
\end{compactenum}
For both tasks, we use PaSST (Patchout Spectrogram Transformer) \cite{Koutini:2022:PaSST} as a pre-trained feature extractor, specifically the \texttt{hear21passt} package \cite{Koutini:2022:hear21passt}, which provides AudioSet-pretrained checkpoints following the HEAR 2021 NeurIPS Challenge API. For the feature extraction, all audio is resampled to \unit[32]{kHz} and zero-padded to a fixed length of \unit[5]{s} when shorter than \unit[5]{s}. The frozen PaSST backbone provides 768-dimensional clip-level embeddings by averaging timestamp-level representations. These embeddings are the input to a linear probe classifier (9-way or 73-way) with class-weighted cross-entropy loss using the AdamW optimizer; each class weight is set to $N / (K \cdot n_k)$, with $N$ the total number of training samples, $K$ the number of classes, and $n_k$ the count of class $k$. We monitor validation accuracy for early stopping (patience: 10 epochs) and select the best checkpoint based on validation performance. The model is then evaluated on the held-out test set.

\begin{figure*}[!tbh]
  \centering
  \includegraphics[width=\textwidth, trim=0.3cm 0.2cm 0.3cm 0.2cm, clip]{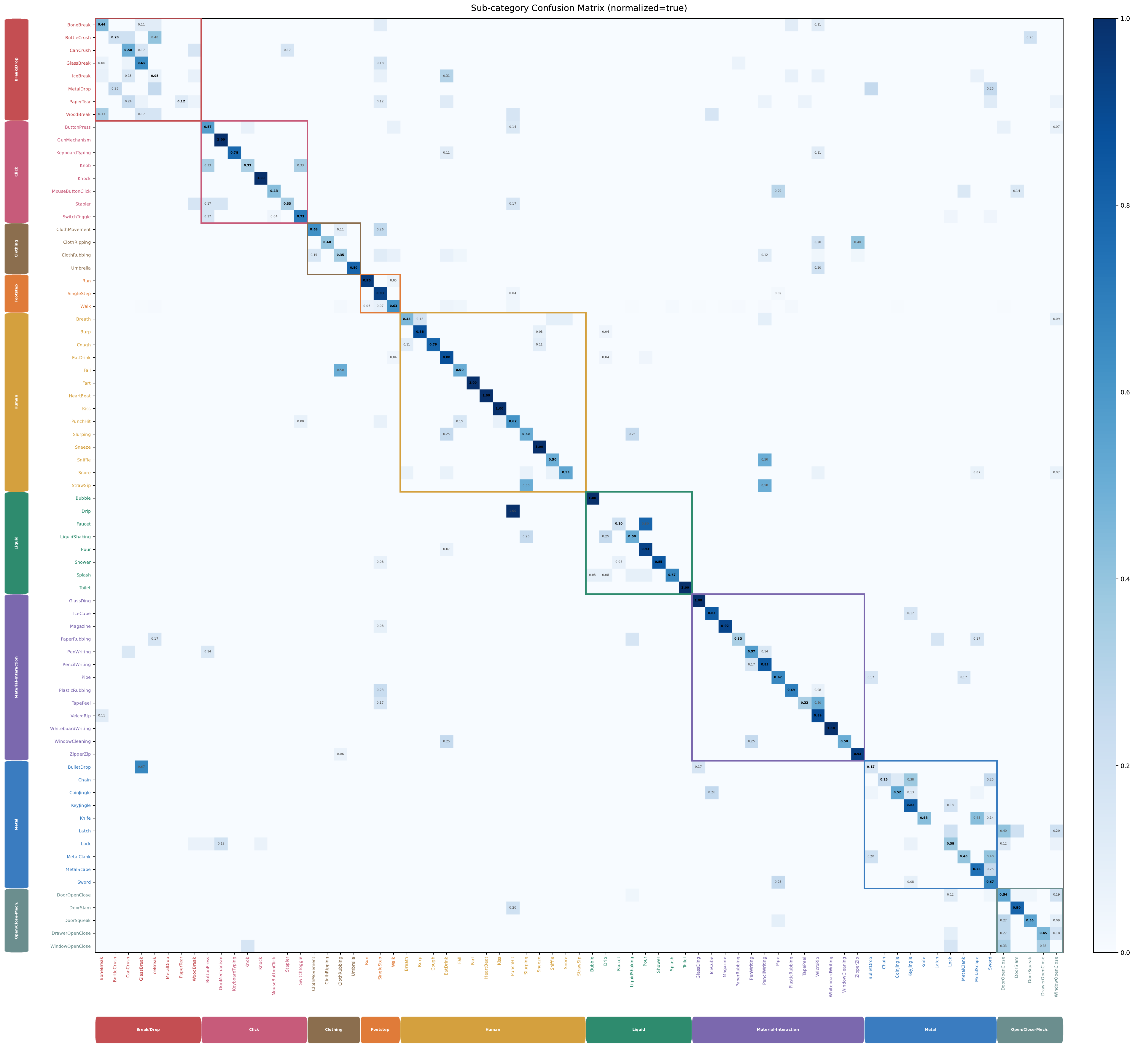}
  \caption{Confusion matrix on the sub-category test set of FoleySet.}
  \label{fig:confmat2}
  \vspace*{-4mm}
\end{figure*}

We evaluate performance using overall accuracy (micro-avg) and F1 scores (macro-averaged and weighted). Both tasks are evaluated on the same held-out test set ($N=1000$). Overall results are listed in Table~\ref{tab:cls_overall}, with per major-class breakdowns in Table~\ref{tab:cls_perclass} and the corresponding confusion matrix in Fig.~\ref{fig:confmat} and Fig.~\ref{fig:confmat2}.

\subsection{Major-Category Classification}
\begin{figure}
  \centering
  \hspace*{-12mm}
  \includegraphics[width=\columnwidth]{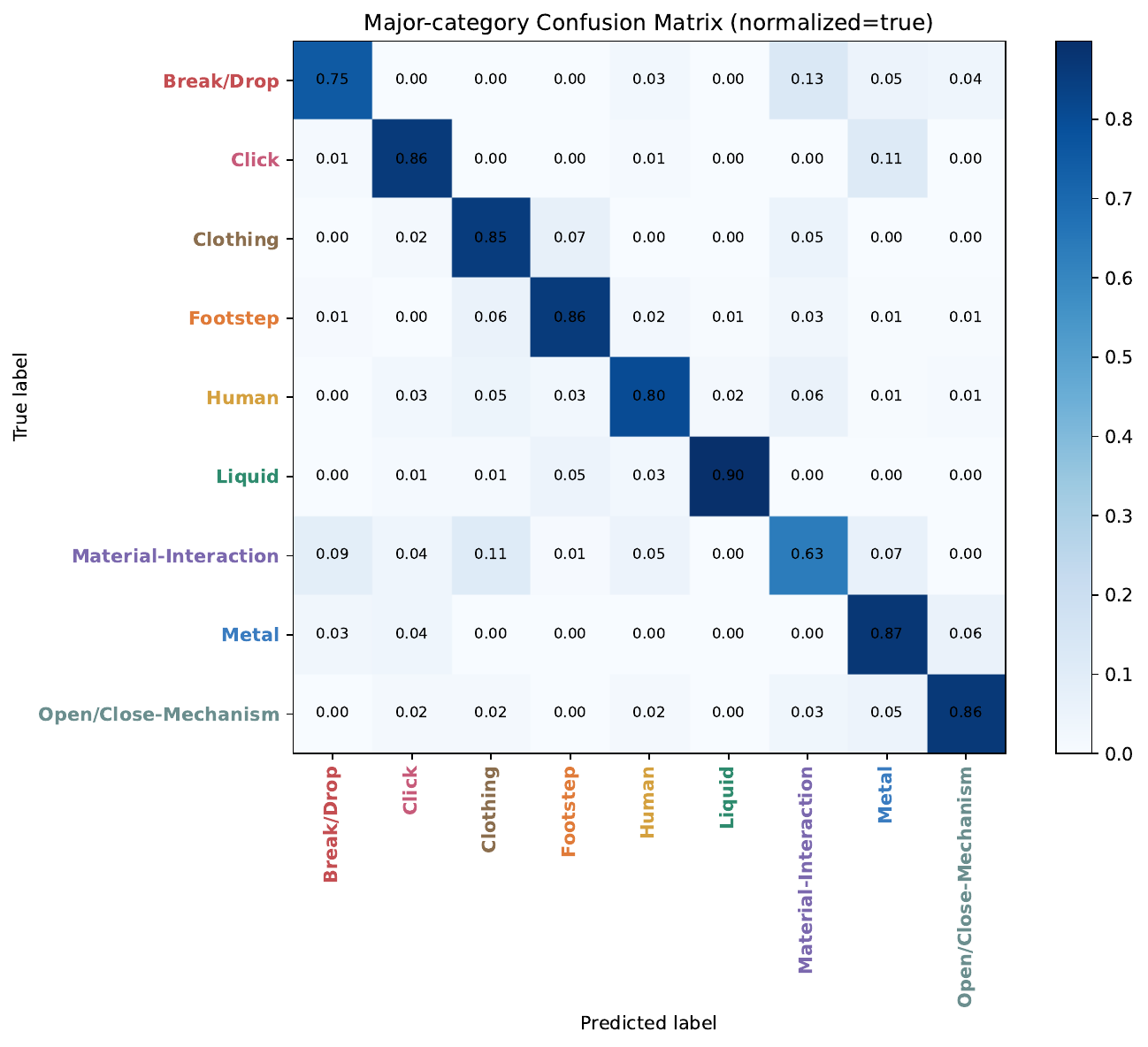}
  \caption{Confusion matrix on the major-category test set of FoleySet.}
  \label{fig:confmat}
\end{figure}

On the 9-way major-category test set, the baseline achieves overall 0.82 accuracy, 0.80 macro-F1, and 0.83 weighted-F1. Per-class results show strong performance for categories with distinctive acoustic signatures (e.g., \texttt{Footstep}: F1=0.90; \texttt{Liquid}: F1=0.92), while comparatively lower F1 scores are observed for {\ttfamily Material-\hspace{0pt}Interaction} (F1=0.65) and \texttt{Clothing} (F1=0.68), suggesting that these categories may be more difficult to classify because they are acoustically less homogeneous and exhibit broader intra-class variation. The row-normalized confusion matrix (Fig.~\ref{fig:confmat}) further shows strong diagonal dominance overall, with \texttt{Liquid} achieving the highest class-conditional accuracy (0.90) and {\ttfamily Material-\hspace{0pt}Interaction} the lowest (0.63). Most \texttt{Break/Drop} errors are predicted as {\ttfamily Material-\hspace{0pt}Interaction} (0.13), reflecting similar transient impact cues. {\ttfamily Material-\hspace{0pt}Inter} \texttt{action} shows the broadest confusion, most often mapped to \texttt{Clothing} (0.11), \texttt{Break/Drop} (0.09), and \texttt{Metal} (0.07). Near-symmetric confusions can also be observed  between \texttt{Clothing} and \texttt{Footstep} (0.07/0.06) and between \texttt{Metal} and {\ttfamily Open/Close-\hspace{0pt}Mechanism} (0.06/0.05).

\subsection{Sub-Category Classification}
For Task~2, performance decreases to 0.64 accuracy, 0.56 macro-F1, and 0.64 weighted-F1 on the 73-way sub-category test set (full per-class results are provided in the Appendix in Table~\ref{tab:sub_results_all}). This unsurprising drop relative to Exp.~1 reflects the substantially greater difficulty of fine-grained Foley recognition with many classes. Several sub-categories with acoustically distinctive characteristics still achieve strong results. For example, \texttt{Fart}, \texttt{HeartBeat}, \texttt{Toilet}, and \texttt{WhiteboardWriting} all achieve an F1 score of 1.00, while \texttt{Kiss} (F1 = 0.96), \texttt{Knock} (F1 = 0.96), and \texttt{Bubble} (F1 = 0.91) also perform well. In contrast, several sub-categories are particularly difficult to recognize, including zero-recall classes such as \texttt{Drip} (support = 5), \texttt{Latch} (5), \texttt{MetalDrop} (4), \texttt{StrawSip} (2), \texttt{WindowOpenClose} (6), and \texttt{WoodBreak} (6), as well as low-performing classes such as \texttt{IceBreak} (support = 13, F1 = 0.09) and \texttt{PaperTear} (support = 17, F1 = 0.21). Taken together, these cases suggest that weak sub-category performance is influenced by both limited learning resources and strong acoustic overlap between related classes. This interpretation is visualized by the sub-category confusion matrix Fig.~\ref{fig:confmat2}, where errors often cluster within acoustically related regions rather than being distributed uniformly across the taxonomy. For example, \texttt{IceBreak} shows only a small recall (about 0.08), with strong confusion with nearby break-related classes, while \texttt{PaperTear} shows a similarly weak recall (about 0.12) together with visible off-diagonal confusion toward related tearing-like classes. A similar pattern appears among mechanism-related categories. For example, \texttt{DoorOpenClose} (R = 0.54), \texttt{DrawerOpenClose} (R = 0.45), and \texttt{WindowOpenClose} (R = 0.00) are often confused with other nearby mechanism classes in the confusion matrix, suggesting that the model captures the broader open/close sound family more reliably than the fine-grained distinctions between specific sources.
We also observe cases of high precision but low recall, such as Chain (P = 1.00, R = 0.25) and \texttt{PaperTear} (P = 1.00, R = 0.12), indicating that the classifier is overly conservative for some rare classes: it correctly identifies a small number of highly prototypical examples, but assigns the majority of instances to acoustically similar neighboring categories. 

Overall, these results provide a reproducible baseline for major and sub-category recognition under our taxonomy, while highlighting that Foley sub-category classification is challenged not only by limited learning resources, but also by fine-grained acoustic overlap and the difficulty of separating closely related classes with broad intra-class variation.

\section{Conclusion}
We presented FoleySet, a novel human-annotated Foley dataset built upon a consistent and data-informed taxonomy. The work has two main contributions.
First, it addresses the need for a Foley-related dataset in an important yet underexplored subdomain of the broader field of sound effects, providing a resource that facilitates future Foley research while also offering value for other data-driven audio tasks. Through workflow-derived annotations and carefully curated audio clips, FoleySet is intended to serve as a practical resource for developing Foley and broader sound-effect models with potential relevance to real-world production workflows.
Second, it introduces a novel Foley-specific taxonomy with a systematic and well-documented design process, rationale, and structure, providing both a basis for future maintenance, refinement, adaptation, and extension and a reference for other researchers developing taxonomies for specialized audio domains. While the current release provides a strong foundation, its annotations were produced by a single annotator and its subcategory distribution remains long-tailed. Building on this work, future efforts will explore independent annotation validation, broader coverage of underrepresented categories, and evaluation across additional Foley-related tasks.

\FloatBarrier
\bibliographystyle{IEEEtranDAFx}
\bibliography{DAFx26_tmpl} 

@article{fonseca2022fsd50k,
  title   = {{FSD50K}: An Open Dataset of Human-Labeled Sound Events},
  author  = {Fonseca, Eduardo and Favory, Xavier and Pons, Jordi and
             Font, Frederic and Serra, Xavier},
  journal = {{IEEE/ACM} Trans. on Audio, Speech, and Language Processing},
  volume  = {30},
  pages   = {829--852},
  year    = {2022}
}

@inproceedings{piczak2015dataset,
  title = {{ESC}: {Dataset} for {Environmental Sound Classification}},
  author = {Piczak, Karol J.},
  booktitle = {Proc. of the 23rd {ACM Conf.} on {Multimedia} (MM)},
  date = {2015-10-13},
  year = {2015},
  url = {http://dl.acm.org/citation.cfm?doid=2733373.2806390},
  doi = {10.1145/2733373.2806390},
  location = {{Brisbane, Australia}},
}

@inproceedings{salamon2014urbansound8k,
  title        = {A Dataset and Taxonomy for Urban Sound Research},
  author       = {Salamon, Justin and Jacoby, Christopher and Bello, Juan Pablo},
  booktitle    = {Proc. of the 22nd ACM Intl. Conf. on Multimedia (MM)},
  year         = {2014},
  address      = {Orlando, FL},
  doi          = {10.1145/2647868.2655045}
}

@article{Lewis:2015:VentriloquialActs,
    Author  = {M. Lewis},
    Title   = {Ventriloquial Acts: Critical Reflections on the Art of Foley},
    Journal = {The New Soundtrack},
    Volume  = 5,
    Number  = 2,
    Year    = 2015,
    Pages   = {103--120},
    doi     = {10.3366/sound.2015.0073}
}

@inproceedings{Trento:2011:FoleySoundsVsRealSounds,
Author    = {Stefano Trento and Amalia de G{\"o}tzen},
Title     = {Foley Sounds vs. Real Sounds},
Booktitle = {Proc. Sound and Music Computing Conference (SMC)},
Year      = {2011}
}

@article{Donaldson:2014:InvisibleBody,
  author  = {Lucy Fife Donaldson},
  title   = {The Work of an Invisible Body: The Contribution of Foley Artists to On-Screen Effort},
  journal = {Alphaville: Journal of Film and Screen Media},
  number  = {7},
  year    = {2014}
}

@inproceedings{Chung:2024:TFoley,
  author    = {Chung, Yoonjin and Lee, Junwon and Nam, Juhan},
  title     = {{T-FOLEY}: A Controllable Waveform-Domain Diffusion Model
               for Temporal-Event-Guided Foley Sound Synthesis},
  booktitle = {Proc. IEEE Int. Conf. Acoustics, Speech, and Signal
               Processing (ICASSP)},
  address   = {Seoul, Korea},
  year      = {2024}
}

@inproceedings{Colombo:2025:MambaFoley,
Author    = {M. F. Colombo and F. Ronchini and L. Comanducci and F. Antonacci},
Title     = {{MambaFoley}: Foley Sound Generation using Selective State-Space Models},
Booktitle = {Proc. IEEE International Conference on Acoustics, Speech and Signal Processing (ICASSP)},
Address   = {Hyderabad, India},
Year      = {2025},
Pages     = {1--5},
Doi       = {10.1109/ICASSP49660.2025.10888864}
}

@electronic{Freesound,
    Author = {{Freesound}},
    Title = {{Freesound}},
    Howpublished = {{A}ccessed: Sep. 2025.},
    note = {{A}vailable: \href{https://freesound.org}{https://freesound.org}}}

@electronic{BBCSoundEffects,
    Author = {{BBC}},
    Title = {{BBC Sound Effects}},
    Howpublished = {{A}ccessed: Sep.--Nov. 2025.},
    note = {{A}vailable: \href{https://sound-effects.bbcrewind.co.uk/}{https://sound-effects.bbcrewind.co.uk/}}}

@misc{ucs_category_list2024,
  author       = {Nielsen, Tim and Drury, Justin and Paquin, Kai and others},
  title        = {{Universal Category System (UCS)} Category List},
  year         = {2024},
  howpublished = {\url{https://universalcategorysystem.com/}},
  note         = {Version 8.2.1, accessed March 10, 2026}
}

@inproceedings{Koutini:2022:PaSST,
Author    = {Khaled Koutini and Jan Schl{\"u}ter and Hamid Eghbal{-}zadeh and Gerhard Widmer},
Title     = {Efficient Training of Audio Transformers with Patchout},
Booktitle = {Proc. Interspeech},
Year      = {2022},
Pages     = {2753--2757},
Doi       = {10.21437/Interspeech.2022-227}
}

@InProceedings{Gemmeke:2017:AudioSet,
    Author =    {J. F. Gemmeke and D. P. W. Ellis and D. Freedman and A. Jansen and W. Lawrence and R. C. Moore and M. Plakal and M. Ritter},
    Title =     {Audio set: An ontology and human-labeled dataset for audio events},
    Booktitle = {Proc. IEEE Intl. Conf. Acoustics, Speech, and Signal Processing (ICASSP)},
    Year =      {2017}}

@article{Ghose:2020:AutoFoley,
  author={Ghose, Sanchita and Prevost, John Jeffrey},
  journal={IEEE Trans. on Multimedia}, 
  title={AutoFoley: Artificial Synthesis of Synchronized Sound Tracks for Silent Videos With Deep Learning}, 
  year={2021},
  volume={23},
  number={},
  pages={1895-1907},
  doi={10.1109/TMM.2020.3005033}}

@InProceedings{Luo:2023:DiffFoley,
    Author =    {S. Luo and C. Yan and C. Hu and H. Zhao},
    Title =     {Diff-{Foley}: Synchronized Video-to-Audio Synthesis with Latent Diffusion Models},
    Booktitle = {Advances in Neural Information Processing Systems 36 (NeurIPS)},
    Year =      {2023},
    Address =   {New Orleans, LA},
    Month =     {Dec. 10--16,}}

@InProceedings{Choi:2023:FoleySynthesis,
    Author = 	 {K. Choi and J. Im and L. M. Heller and B. McFee and K. Imoto and Y. Okamoto and M. Lagrange and S. Takamichi},
    Title = 	 {Foley Sound Synthesis at the DCASE 2023 Challenge},
    Booktitle = 	 {Proc. Workshop Detection and Classification of Acoustic Scenes and Events (DCASE)},
    Year = 	 {2023},
    Address = 	 {Tampere, Finland},
    Month = 	 {Sept. 21--22,}}

@misc{soundideas_hollywood_foley_fx,
  author       = {{Sound Ideas}},
  title        = {Hollywood Foley FX},
  howpublished = {Sound Ideas product page},
  url          = {https://sound-ideas.com/products/hollywood-foley-fx},
  note         = {Accessed: Sep. 2025},
  year         = {2025}
}

@misc{soundideas_foley_sound_effects,
  author       = {{Sound Ideas}},
  title        = {Foley Sound Effects},
  howpublished = {Sound Ideas product page},
  url          = {https://sound-ideas.com/products/foley-sound-effects},
  note         = {Accessed: Sep. 2025},
  year         = {2025}
}

@misc{soundideas_art_of_foley,
  author       = {{Sound Ideas}},
  title        = {Art of Foley Sound Effects Library},
  howpublished = {Sound Ideas product page},
  url          = {https://sound-ideas.com/products/art-of-foley-sound-effects-library},
  note         = {Accessed: Sep. 2025},
  year         = {2025}
}

@misc{soundideas_hd_foley,
  author       = {{Sound Ideas}},
  title        = {HD -- Foley Sound Effects},
  howpublished = {Sound Ideas product page},
  url          = {https://sound-ideas.com/products/hd-foley-sound-effects},
  note         = {Accessed: Sep. 2025},
  year         = {2025}
}

@misc{soundideas_ultimate_foley_collection,
  author       = {{Sound Ideas}},
  title        = {Ultimate Foley Sound Effects Collection},
  howpublished = {Sound Ideas product page},
  url          = {https://sound-ideas.com/products/ultimate-foley-sound-effects-collection},
  note         = {Accessed: Sep. 2025},
  year         = {2025}
}

@misc{soundideas_fight_foley,
  author       = {{Sound Ideas}},
  title        = {Fight Foley Sound Effects},
  howpublished = {Sound Ideas product page},
  url          = {https://sound-ideas.com/products/fight-foley-sound-effects},
  note         = {Accessed: Sep. 2025},
  year         = {2025}
}

@misc{soundideas_foley_footsteps,
  author       = {{Sound Ideas}},
  title        = {Foley Footsteps Sound Effects Library},
  howpublished = {Sound Ideas product page},
  url          = {https://sound-ideas.com/products/foley-footsteps-sound-effects-library},
  note         = {Accessed: Sep. 2025},
  year         = {2025}
}

@misc{Koutini:2022:hear21passt,
  author       = {Khaled Koutini},
  title        = {hear21passt: {PaSST} pre-trained models},
  year         = {2022},
  howpublished = {\url{https://github.com/kkoutini/PaSST}},
  note         = {Accessed: Feb. 2, 2026},
}

@inproceedings{Ma:2022:RepLearning,
  Author    = {A. B. Ma and A. Lerch},
  Title     = {Representation learning for the automatic indexing of sound effects libraries},
  Booktitle = {Proc. 23rd Int. Society for Music Information Retrieval Conf. (ISMIR)},
  Year      = {2022},
  Address   = {Bengaluru, India},
  Pages     = {387--394}}

@inproceedings{Peeters:2020:SoundClassPostProd,
  Author    = {G. G. Peeters and J. D. Reiss},
  Title     = {A deep learning approach to sound classification for film audio post-production},
  Booktitle = {Proc. 148th AES Convention},
  Year      = {2020},
  Number    = {10322}}

@inproceedings{Moffat:2017:UnsupTaxonomy,
  Author    = {D. Moffat and D. Ronan and J. D. Reiss},
  Title     = {Unsupervised taxonomy of sound effects},
  Booktitle = {Proc. 20th Intl. Conf. on Digital Audio Effects (DAFx)},
  Year      = {2017},
  Address   = {Edinburgh, UK}}

@techreport{EBU:2011:R128,
  author      = {{European Broadcasting Union}},
  title       = {{EBU} {R} 128: Loudness Normalisation and
                 Permitted Maximum Level of Audio Signals},
  institution = {European Broadcasting Union (EBU)},
  type        = {Tech.\ Rep.},
  number      = {R 128},
  address     = {Geneva, Switzerland},
  year        = {2011}
}

@InProceedings{Lin:2023:Soundify,
    Author =    {D. C.-E. Lin and A. Germanidis and C. Valenzuela and Y. Shi and N. Martelaro},
    Title =     {Soundify: Matching Sound Effects to Video},
    Booktitle = {Proc. ACM Symp. User Interface Software and Technology (UIST)},
    Year =      {2023},
    Address =   {San Francisco, CA},
    Month =     {Oct. 29--Nov. 1,}
}

@InProceedings{Dixit:2026:FoleyBench,
    Author    = {S. Dixit and K. Saito and Z. Zhong and Y. Mitsufuji and C. Donahue},
    Title     = {FoleyBench: A Benchmark for Video-to-Audio Models},
    Booktitle = {Proc. {IEEE} Intl. Conf. on Acoustics, Speech and Signal Processing (ICASSP)},
    Year      = {2026},
    Address   = {Barcelona, Spain},
    Month     = {May 4--8},
}

@Article{Fu:2024:MINT,
    Author  = {R. Fu and S. Shi and H. Guo and T. Wang and C. Qiang and Z. Wen and J. Tao and X. Qi and Y. Lu and X. Wang and Z. Wang and Y. Liu and X. Liu and S. Zhang and G. Li},
    Title   = {MINT: a Multi-modal Image and Narrative Text Dubbing Dataset for Foley Audio Content Planning and Generation},
    Journal = {arXiv preprint arXiv:2406.10591},
    Year    = {2024}
}

@Article{Garcia:2025:6KSFx,
    Author  = {N. Garcia and J. Reiss},
    Title   = {6KSFx Synth Dataset},
    Journal = {arXiv preprint arXiv:2501.17198},
    Year    = {2025}
}

\begin{table*}[!t]
\centering
\section{Appendix}
\vspace{0.5em}
\small
\setlength{\tabcolsep}{4pt}
\renewcommand{\arraystretch}{1.12}
\caption{Keyword-to-category mapping for the Foley taxonomy.}
\label{tab:kw_sub_major_2col_vertical}
\begin{tabularx}{\textwidth}{
    X >{\raggedright\arraybackslash}p{2.5cm} >{\raggedright\arraybackslash}p{2.7cm} |
    X >{\raggedright\arraybackslash}p{2.5cm} >{\raggedright\arraybackslash}p{2.7cm}}
\toprule
\textbf{Keywords} & \textbf{Sub-category} & \textbf{Major category} &
\textbf{Keywords} & \textbf{Sub-category} & \textbf{Major category} \\
\midrule
walk; move; gravel; floor & Walk & Footstep &
tape; peel & TapePeel & Material-Interaction \\
footstep; feet; foot; shoe; boot & SingleStep & Footstep &
writing; board & WhiteboardWriting & Material-Interaction \\
run & Run & Footstep &
window; cleaning; wipe & WindowCleaning & Material-Interaction \\
burp; gurgle & Burp & Human &
switch; lamp & SwitchToggle & Click \\
eat; drink & EatDrink & Human &
push & ButtonPress & Click \\
cough & Cough & Human &
tap & Knock & Click \\
snore; groaning & Snore & Human &
keyboard; typing; typewriter & KeyboardTyping & Click \\
hit; punch & PunchHit & Human &
mouse; click & MouseButtonClick & Click \\
breath & Breath & Human &
stapler; card & Stapler & Click \\
kiss & Kiss & Human &
knob & Knob & Click \\
fart; gusts & Fart & Human &
gun & GunMechanism & Click \\
gasping & Sneeze & Human &
pour; water & Pour & Liquid \\
fall & Fall & Human &
shower & Shower & Liquid \\
sucking & Slurping & Human &
splash & Splash & Liquid \\
sniff & Sniffle & Human &
toilet; flush & Toilet & Liquid \\
knock & HeartBeat & Human &
drip & Drip & Liquid \\
gulping & StrawSip & Human &
bubble; boiling & Bubble & Liquid \\
coin; pocket & CoinJingle & Metal &
flow & Faucet & Liquid \\
lock & Lock & Metal &
shake; slosh; fluid & LiquidShaking & Liquid \\
sword & Sword & Metal &
crack; smash & GlassBreak & Break/Drop \\
key; jingle & KeyJingle & Metal &
tear & PaperTear & Break/Drop \\
chain & Chain & Metal &
break & IceBreak & Break/Drop \\
knife & Knife & Metal &
bone & BoneBreak & Break/Drop \\
pipe & Pipe & Metal &
wood; branch & WoodBreak & Break/Drop \\
bullet; drop & BulletDrop & Metal &
can; crush & CanCrush & Break/Drop \\
latch & Latch & Metal &
bolt & MetalDrop & Break/Drop \\
clank & MetalClank & Metal &
bottle; broken & BottleCrush & Break/Drop \\
brass; scratch & MetalScape & Metal &
door; closed & DoorOpenClose & Open/Close-Mechanism \\
zipper; zip; unzip & ZipperZip & Material-Interaction &
drawer & DrawerOpenClose & Open/Close-Mechanism \\
pencil & PencilWriting & Material-Interaction &
squeak & DoorSqueak & Open/Close-Mechanism \\
plastic; rub & PlasticRubbing & Material-Interaction &
open & WindowOpenClose & Open/Close-Mechanism \\
magazine; book & Magazine & Material-Interaction &
slam & DoorSlam & Open/Close-Mechanism \\
velcro; rip; sticky & VelcroRip & Material-Interaction &
cloth; leather; fur; pants & ClothRubbing & Clothing \\
glass; clink & GlassDing & Material-Interaction &
rustle; dress & ClothMovement & Clothing \\
pen & PenWriting & Material-Interaction &
fabric & ClothRipping & Clothing \\
paper & PaperRubbing & Material-Interaction &
umbrella & Umbrella & Clothing \\
ice & IceCube & Material-Interaction &
 &  &  \\
\bottomrule
\end{tabularx}
\end{table*}

\begin{table*}[p]
\centering
\small
\setlength{\tabcolsep}{4pt}
\renewcommand{\arraystretch}{1.12}
\caption{Per-class results for 73-way sub-category classification on the FoleySet test set (N=1000). P = Precision, R = Recall, Sup = Support.}
\label{tab:sub_results_all}
\begin{tabular*}{\textwidth}{@{\extracolsep{\fill}}lcccc|lcccc}
\toprule
\textbf{Class} & \textbf{P} & \textbf{R} & \textbf{F1} & \textbf{Sup} & \textbf{Class} & \textbf{P} & \textbf{R} & \textbf{F1} & \textbf{Sup} \\
\midrule
BoneBreak        & 0.44 & 0.44 & 0.44 &   9 & LiquidShaking     & 0.29 & 0.50 & 0.36 &   4 \\
BottleCrush      & 0.50 & 0.20 & 0.29 &   5 & Lock              & 0.40 & 0.38 & 0.39 &  16 \\
Breath           & 0.62 & 0.45 & 0.53 &  11 & Magazine          & 0.92 & 0.92 & 0.92 &  12 \\
Bubble           & 0.83 & 1.00 & 0.91 &   5 & MetalClank        & 0.50 & 0.40 & 0.44 &   5 \\
BulletDrop       & 0.20 & 0.17 & 0.18 &   6 & MetalDrop         & 0.00 & 0.00 & 0.00 &   4 \\
Burp             & 0.91 & 0.88 & 0.89 &  24 & MetalScape        & 0.30 & 0.75 & 0.43 &   4 \\
ButtonPress      & 0.50 & 0.57 & 0.53 &  14 & MouseButtonClick  & 0.75 & 0.43 & 0.55 &   7 \\
CanCrush         & 0.25 & 0.50 & 0.33 &   6 & PaperRubbing      & 0.40 & 0.33 & 0.36 &   6 \\
Chain            & 1.00 & 0.25 & 0.40 &   8 & PaperTear         & 1.00 & 0.12 & 0.21 &  17 \\
ClothMovement    & 0.75 & 0.63 & 0.69 &  19 & PenWriting        & 0.57 & 0.57 & 0.57 &   7 \\
ClothRipping     & 1.00 & 0.40 & 0.57 &   5 & PencilWriting     & 0.48 & 0.83 & 0.61 &  12 \\
ClothRubbing     & 0.47 & 0.35 & 0.40 &  26 & Pipe              & 0.36 & 0.67 & 0.47 &   6 \\
CoinJingle       & 0.86 & 0.52 & 0.65 &  23 & PlasticRubbing    & 0.56 & 0.69 & 0.62 &  13 \\
Cough            & 1.00 & 0.79 & 0.88 &  19 & Pour              & 0.81 & 0.93 & 0.87 &  28 \\
DoorOpenClose    & 0.50 & 0.54 & 0.52 &  26 & PunchHit          & 0.28 & 0.62 & 0.38 &  13 \\
DoorSlam         & 0.57 & 0.80 & 0.67 &   5 & Run               & 0.59 & 0.95 & 0.73 &  20 \\
DoorSqueak       & 0.75 & 0.55 & 0.63 &  11 & Shower            & 1.00 & 0.85 & 0.92 &  13 \\
DrawerOpenClose  & 0.71 & 0.45 & 0.56 &  11 & SingleStep        & 0.52 & 0.93 & 0.67 &  45 \\
Drip             & 0.00 & 0.00 & 0.00 &   5 & Slurping          & 0.40 & 0.50 & 0.44 &   4 \\
EatDrink         & 0.46 & 0.88 & 0.60 &  24 & Sneeze            & 0.69 & 1.00 & 0.82 &   9 \\
Fall             & 0.14 & 0.50 & 0.22 &   4 & Sniffle           & 0.50 & 0.50 & 0.50 &   4 \\
Fart             & 1.00 & 1.00 & 1.00 &   9 & Snore             & 0.89 & 0.53 & 0.67 &  15 \\
Faucet           & 0.50 & 0.20 & 0.29 &   5 & Splash            & 0.73 & 0.67 & 0.70 &  12 \\
GlassBreak       & 0.55 & 0.65 & 0.59 &  17 & Stapler           & 0.67 & 0.33 & 0.44 &   6 \\
GlassDing        & 0.78 & 1.00 & 0.88 &   7 & StrawSip          & 0.00 & 0.00 & 0.00 &   2 \\
GunMechanism     & 0.43 & 1.00 & 0.60 &   3 & SwitchToggle      & 0.89 & 0.71 & 0.79 &  24 \\
HeartBeat        & 1.00 & 1.00 & 1.00 &   2 & Sword             & 0.42 & 0.67 & 0.52 &  12 \\
IceBreak         & 0.11 & 0.08 & 0.09 &  13 & TapePeel          & 0.67 & 0.33 & 0.44 &   6 \\
IceCube          & 0.42 & 0.83 & 0.56 &   6 & Toilet            & 1.00 & 1.00 & 1.00 &   7 \\
KeyJingle        & 0.50 & 0.82 & 0.62 &  11 & Umbrella          & 0.80 & 0.80 & 0.80 &   5 \\
KeyboardTyping   & 1.00 & 0.78 & 0.88 &   9 & VelcroRip         & 0.42 & 0.89 & 0.57 &   9 \\
Kiss             & 0.92 & 1.00 & 0.96 &  11 & Walk              & 0.97 & 0.63 & 0.76 & 231 \\
Knife            & 1.00 & 0.43 & 0.60 &   7 & WhiteboardWriting & 1.00 & 1.00 & 1.00 &   3 \\
Knob             & 0.33 & 0.33 & 0.33 &   3 & WindowCleaning    & 1.00 & 0.50 & 0.67 &   4 \\
Knock            & 0.93 & 1.00 & 0.96 &  13 & WindowOpenClose   & 0.00 & 0.00 & 0.00 &   6 \\
Latch            & 0.00 & 0.00 & 0.00 &   5 & WoodBreak         & 0.00 & 0.00 & 0.00 &   6 \\
                 &      &      &      &     & ZipperZip         & 0.83 & 0.94 & 0.88 &  16 \\
\midrule
\multicolumn{5}{l}{Accuracy = 0.64 (N = 1000)} & Macro avg & 0.60 & 0.59 & 0.56 & 1000 \\
\multicolumn{5}{l}{}                           & Wtd avg   & 0.71 & 0.64 & 0.64 & 1000 \\
\bottomrule
\end{tabular*}
\end{table*}

\end{document}